\documentclass[11pt]{iopart}

\usepackage{amstext}

\usepackage{iopams}
\usepackage{graphics}
\usepackage{graphicx}

\usepackage{epsfig}
\newcommand{\dd}{\text{d}}
\newcommand{\sR}{\text{\tiny R}}
\newcommand{\ee}{\text{e}}

\newcommand{\p}{\partial}
\newcommand{\bx}{\text{\bf x}}
\newcommand{\by}{\text{\bf y}}

\newcommand{\bj}{\text{\bf j}}

\newcommand{\bq}{\text{\bf q}}

\newcommand{\bu}{\text{\bf u}}

\begin{document}


\title[Current distribution in systems with anomalous diffusion: RG approach]
	{Current distribution in systems with anomalous diffusion: 
	renormalisation group approach}

\author{Vivien Lecomte$^{1,2}$, Uwe C. T\"auber$^{3,1}$, 
	Fr\'ed\'eric van Wijland$^{1,2}$}

\address{\ $^1$ Laboratoire de Physique Th\'eorique (CNRS UMR 8627), \\
	Universit\'e de Paris-Sud, 91405 Orsay cedex, France}

\address{\ $^2$ Laboratoire Mati\`ere et Syst\`emes Complexes 
	(CNRS UMR 7057), \\ Universit\'e Paris VII -- Denis Diderot, 
        10 rue Alice Domon et L\'eonie Duquet, \\ 75205 Paris cedex 13, France}

\address{\ $^3$ Department of Physics and Center for Stochastic Processes 
	in Science and Engineering, Virginia Polytechnic Institute and State 
	University, \\ Blacksburg, VA 24061-0435, USA}

\eads{\mailto{tauber@vt.edu}, \mailto{Frederic.Van-Wijland@th.u-psud.fr}}

\begin{abstract}
We investigate the asymptotic properties of the large deviation function of 
the integrated particle current in systems, in or out of thermal equilibrium, 
whose dynamics exhibits anomalous diffusion. 
The physical systems covered by our study include mutually repelling particles
with a drift, a driven lattice gas displaying a continuous nonequilibrium phase
transition, and particles diffusing in a anisotropic random advective field.
It is exemplified how renormalisation group techniques allow for a systematic 
determination of power laws in the corresponding current large deviation 
functions.
We show that the latter are governed by known universal scaling exponents,
specifically, the anomalous dimension of the noise correlators.
\end{abstract}





\section{Introduction}

\subsection{Motivations}
\label{subsec:motivations}

Recently an important concept for the analysis of systems driven out of thermal
equilibrium, namely that of {\em large deviations} in the associated 
probability distributions, has again found considerable attention, both at the 
experimental and theoretical level.
Mathematically, large deviation functions have been defined and introduced long
ago, but only in the early seventies were they shown to play a central role in 
characterising the properties of non-equilibrium steady state (NESS) 
measures, see \cite{varadhan88} for a review.
For some time this idea remained mostly confined within the mathematical 
community until it was exploited in physics at the theoretical level by Evans, 
Cohen, and Morriss who showed numerically that the temporal large deviation 
function of a certain observable, namely the shear stress, in the NESS of 
viscous gases under shear displayed a surprisingly simple symmetry property 
\cite{evanscohenmorriss}.
This symmetry property, now known as the {\em fluctuation relation}, was then
formalised into a theorem by Gallavotti and Cohen \cite{gallavotticohen}. 
Over the past years, a remarkable amount of research has been devoted to the
investigation of the fluctuation relation in detail, with the goal to 
explicitly determine the large deviation function at the core of the theorem's 
formulation \cite{evanssearles94, jarzynski, kurchan, lebowitzspohn}.
It must be emphasised that at about the same period, experimentalists also 
started to employ, as a tool for analysing their data, the concept of large 
deviations \cite{ldf-exp}. 
Their initial interest was grounded in the belief that, much in the same way as
the intensive free energy per degree of freedom constitutes a central quantity 
if one wishes to relate two systems in thermal equilibrium, large deviation 
functions might similarly allow for a comparison of NESS {\em dynamical} 
properties, yet be applicable to systems both in and out of equilibrium. 
Indeed, it has been a highly desirable, but as yet largely elusive 
goal of non-equilibrium statistical mechanics to be able to classify dynamical 
systems (at least in a steady state) according to a restricted number of 
(macro- or mesoscopic) criteria. 
In particular, it was expected that large deviation functions of observables
such as energy flow, particle current (or more generally entropy flow) would 
display a sufficient degree of universality to render this approach fruitful
(see refs. \cite{evanssearles94} through \cite{bertinidesole4}). 

On the theoretical side, a series of recent results however lends support to 
such aspirations.  
By studying systems of interacting and diffusing particles governed by some 
Markovian dynamics, subjected to an external chemical potential gradient, 
Bodineau and Derrida \cite{bodineauderrida1, bodineauderrida2, 
bodineauderrida3}, as well as Bertini, De~Sole, Gabrielli, Jona-Lasinio, and 
Landim \cite{bertinidesole2, bertinidesole3, bertinidesole4}, found a general
strategy to completely determine the large deviation of the particle current 
flowing through the system.
Note, however, that in this setting the applied field (and the related particle
current) become vanishingly small as the system size is increased.  
Then the only two ingredients entering the final expression for the large 
deviation function are the diffusion constant and the equilibrium 
compressibility, which is an astonishingly simple result: 
The large deviations of the current are fully characterised by properties of 
the system at equilibrium or in the linear response regime.

There are two restrictions, however, that apply to this series of recent 
advances. 
Namely, first they all concern systems exhibiting {\em normal} diffusive 
dynamics. 
In physical terms, this means that the continuum limit can be taken by scaling 
time according to $\tau \sim a^2$ proportional to a microscopic length scale 
$a$ squared (often referred to as the hydrodynamic limit). 
There are of course numerous situations in nature where diffusion becomes 
{\em anomalous}, which means that the time and length scales appropriate for 
taking the continuum limit are related by $\tau \sim a^z$ with $z \neq 2$. 
The purpose of the present work is to investigate the particle current large 
deviation functions in such physical systems governed by anomalous diffusion; 
specifically, we shall address the following models: 
(i) mutually excluding particles in a driving `electric' field 
\cite{janssenschmittmann, leungcardy}; here $z = 3/2$ in one space dimension, 
which is equivalent to the scaling for the noisy Burgers equation 
\cite{forsternelsonstephen}, in two dimensions one finds $z_\parallel = 2$ with
logarithmic corrections along the drive direction, while normal diffusive 
behaviour ensues transverse to it, $z_\perp = 2$; 
(ii) critical driven diffusive systems (driven model B) 
\cite{schmittmannzia1, schmittmannzia2}, for which one similarly obtains
$z_\parallel = 12 / (11 - d)$ in dimensions $d < 5$, with again logarithmic 
corrections at the upper critical dimension $d_c = 5$, and $z_\perp = 2$; and
(iii) particle diffusion in a quenched random velocity field, where according 
to the type of disorder, both super- and sub-diffusive behaviour may occur
\cite{honkonen, bouchaudetal}.
Based on a field theory representation of the non-linear Langevin-type 
stochastic differential equations representing the above processes
\cite{janssen1, dedominicis, bauschjanssenwagner, janssen2}, renormalisation 
group (RG) methods will prove instrumental in extracting the appropriate 
asymptotic scaling variables in the long-time and large-system limit, and will
be shown to provide the relevant tool for taking the proper continuum limit 
\cite{schaefer, taeuber}.

The second restriction for most recent investigations is that these studies 
apply only to systems that are weakly out of equilibrium, since, as already
mentioned before, the applied external field scales as the reciprocal of the 
system size. 
To the best of our knowledge, it is only for the one-dimensional exclusion 
process with periodic boundary conditions, with \cite{derridalebowitz} or 
without a bias \cite{lebowitzspohn} in particle propagation, that the current 
fluctuations have been accessed analytically to date. 
The efforts of the present paper therefore bear on systems with anomalous 
diffusion in dimensions $d \geq 1$, some of which driven far away from thermal
equilibrium.

In the course of this work we shall investigate the properties of the current 
large deviation in three different systems, which we define succinctly in the 
following subsection. 
In sections 2, 3, and 4 we present the RG analysis and our results for each of 
these model systems. 
Section~2 itself will also expand on the field-theoretic methods upon which we
shall rely, whereas most of the technical details will be skipped in the
subsequent sections 3 and 4. 
Our conclusions are gathered in section 5.

\subsection{Presentation of the model systems}

We will first investigate the properties of particles (without source nor sink)
subjected to a driving force along one spatial direction (denoted `$\parallel$'
in the following), whose local density $\rho({\bf x},t)$ evolves according to 
the continuity equation 
$\p_t \rho({\bf x},t) = - \boldsymbol{\nabla} \cdot {\bf j} ({\bf x},t)$,
with a particle current ${\bf j} = - D \boldsymbol{\nabla} \rho 
+ \rho \, \bu (\rho) + \boldsymbol{\eta}$ 
constructed from phenomenological considerations.
The first term here is a Fickian diffusion current, the second contribution
originates from the driving force and describes a density-dependent velocity 
field $\bu(\rho)$ along the drive, and finally $\boldsymbol{\eta}$ represents 
Gaussian white noise which accounts for fast degrees of freedom not taken into 
account explicitly in our mesoscopic description.
The former continuum Langevin description arises for instance from
a discrete lattice gas description, which we shall use in the following to
determine the particular form of the velocity field $\bu(\rho)$.
If we transform from a lattice occupation number representation ($n_i = 0,1$) 
of the discretised system (with lattice sites labeled by the index $i$) to 
Ising variables $\phi_i = n_i - 1/2 = \mp 1/2$, particle-hole ($Z_2$) symmetry
dictates that to lowest order in the coarse-grained local density field 
$\phi(\bx,t) = \rho({\bf x},t) - \rho_0$ we have
$u_\parallel(\phi) = \varepsilon (1 - \phi^2) + \ldots$.
Denoting by ${\bf j}_0$ the average current, renaming $\varepsilon = D g / 2$, 
and allowing for a different effective diffusivity $\lambda D$ along the drive 
direction, we thus arrive at the following expressions for the transverse and 
longitudinal (component along the drive) currents
\begin{eqnarray}
  {\bf j}_\perp - {\bf j}_{\perp 0} &=& - D \, \boldsymbol{\nabla}_\perp \phi 
  + \boldsymbol{\eta}_\perp \ , \label{jperp} \\
  j_\parallel - j_{\parallel 0}  &=& - \lambda D \, \nabla_\parallel \phi 
  - \frac{D g}{2} \, \phi^2 + \eta_\parallel \ . \label{jpara}
\end{eqnarray}
The parameter $g$ accounts for the interaction of particles along the driving 
field.
After rescaling such that the transverse current sector is effectively in
equilibrium, the noise correlations read
\begin{eqnarray}
  \langle \eta_{\perp i}({\bf x},t) \, \eta_{\perp j}({\bf y},t') \rangle &=& 
  2 D \, \delta_{ij} \delta^{(d)}(\bx-\by) \delta(t-t') \ , \label{nperp} \\
  \langle \eta_\parallel(\bx,t) \, \eta_\parallel(\by,t') \rangle &=&
  2 D \sigma \, \delta^{(d)}(\bx-\by) \, \delta(t-t') \ . \label{npara}
\end{eqnarray}
Collecting everything into a single stochastic Langevin equation for the
density field $\phi$, we at last arrive at
\begin{eqnarray}
  &&\p_t \phi = D \left( \nabla_\perp^2 + \lambda \, \nabla_\parallel^2 \right)
  \phi + \frac{D g}{2} \ \nabla_\parallel \, \phi^2 + \zeta \ , \label{ddslan} 
  \\ &&\langle \zeta(\bx,t) \, \zeta(\by,t') \rangle 
  = - 2 D \left( \nabla_\perp^2 + \sigma \, \nabla_\parallel^2 \right) 
  \delta^{(d)}(\bx-\by) \, \delta(t-t') \ , \label{ddsnoi}
\end{eqnarray}
where $\zeta = - \boldsymbol{\nabla} \cdot \boldsymbol{\eta}$.
Further motivations for this phenomenological approach, and a detailed study of
the scaling properties of this model can be found in 
\cite{vanbeijerenkutnerspohn, janssenschmittmann, leungcardy, schmittmannzia1,
taeuber}.
We emphasise that in the presence of a drive, and with periodic boundary
conditions, \eref{ddslan} and \eref{ddsnoi} in general describe a 
non-equilibrium system.
In one dimension, they reduce to the noisy Burgers equation 
\cite{forsternelsonstephen}.
\medskip

For our second example, we turn to the standard model of a driven diffusive 
system that exhibits a continuous non-equilibrium phase transition in its 
stationary state \cite{schmittmannzia1, schmittmannzia2}.
In this case, the transverse current looks like the one for the time-dependent 
Ginzburg--Landau model with conserved order parameter, 
\begin{equation} \label{jbperp}
  {\bf j}_\perp - {\bf j}_{\perp 0} = - D \, \nabla_\perp 
  \left( r - \nabla_\perp^2 + \frac{u}{6} \, \phi^2 \right) \phi 
  + \boldsymbol{\eta}_\perp \ ,
\end{equation}
while the current component along the drive direction is still given by
\eref{jpara}, and the noise correlations by \eref{nperp} and \eref{npara}.
For the field $\phi$, this yields the Langevin equation 
\begin{equation} \label{drblan}
  \p_t \phi = D \nabla_\perp^2 \left( r - \nabla_\perp^2 + \frac{u}{6} 
  \, \phi^2 \right) \phi + \lambda D \, \nabla_\parallel^2 \, \phi 
  + \frac{D g}{2} \ \nabla_\parallel \, \phi^2 + \zeta \ , 
\end{equation}
with the noise correlator \eref{ddsnoi}.
The system now has a critical point at $r = 0$ (within the mean-field
approximation), which in the absence of the drive $g = 0$ just describes the
second-order phase transition in the Ising model with Kawasaki dynamics 
(model B).
In the presence of the drive ($g \not= 0$), however, the universality class
changes; in fact, $u$ becomes irrelevant for the critical properties of the
driven model B, which are entirely governed by the non-linear driving term 
\cite{schmittmannzia1}.
\medskip 

Finally, we consider non-interacting particles diffusing in a random velocity 
field, as defined by Honkonen \cite{honkonen}. 
At the mesoscopic level, the system is described over a $d$-dimensional 
continuum by a local density fluctuation $\phi(\bx,t)$ subjected to a random 
field $\psi(\bx_\perp)$ directed along $\text{\bf e}_\parallel$. 
One primary feature of the model is that the amplitude of the field depends 
only on the transverse coordinates $\bx_\perp$.  
The density field $\phi(\bx,t)$ evolves according to the Langevin equation 
$\partial_t \phi + \boldsymbol{\nabla} \cdot \bj =0 $ with a current $\bj$ that
accounts for the random field $\psi(\bx_\perp)$ solely in the parallel 
direction:
\begin{eqnarray} \label{eq:honk_currents}
  \bj_\perp &= - D_\perp \boldsymbol{\nabla}_\perp \phi + 
  \boldsymbol{\eta}_\perp \ , \\
  j_\parallel &= - D_\parallel \, \nabla_\parallel \phi - \psi \, \phi 
  + \eta_\parallel \ .
\end{eqnarray} 
The thermal diffusion of particles is encoded by the white noise 
$\boldsymbol{\eta}$ which enters the expression for the current.
Its correlations ensure that both the transverse and longitudinal sectors are
in equilibrium,
\begin{eqnarray} \label{eq:honk_noise}
  \big\langle \eta_{\perp, i}(\bx,t) \, \eta_{\perp, j}(\bx',t') \big\rangle 
  & = 2 D_\perp \delta_{ij} \delta(\bx'-\bx) \delta(t'-t) \ , \\
  \big\langle \eta_{\parallel}(\bx,t) \, \eta_{\parallel }(\bx',t') \big\rangle
  & = 2 D_\parallel \delta(\bx'-\bx)\delta(t'-t) \ . 
\end{eqnarray}
Depending on the spatial correlations of the random field 
$\psi(\bx_\perp) \, \text{\bf e}_\parallel$, particles can behave either
sub- or superdiffusively \cite{bouchaudetal, honkonen}.
We restrict our analysis to the case of a spatially uncorrelated random field 
with Gaussian distribution, whose correlations are described by a constant 
amplitude $\lambda$:
\begin{eqnarray}
  \big\langle \psi(\bx_\perp,t) \, \psi(\bx'_\perp,t') \big\rangle = 
  \lambda \, \delta(\bx'_\perp-\bx_\perp)\delta(t'-t)
\end{eqnarray}
This has the consequence that particles behave superdiffusively in the
longitudinal direction, with a dynamical exponent $z_\parallel = 4 / (5-d)$ in
dimensions $d < d_c = 3$ \cite{honkonen}.  
In contrast to the previous systems, the superdiffusion here stems from the 
anomalous, but {\em equilibrium} wandering of particles in the quenched random 
velocity field, rather than from interactions that drive the system out of 
equilibrium.

\subsection{Large deviation functions}
\label{subsec:ldf}

In each of these systems the goal will be to determine the probability 
distribution $P(Q,t)$ of the total integrated current in the longitudinal 
direction $Q(t) = \int_0^t \dd t' \int \dd^dx \, j_\parallel(\bx ,t')$ up to 
time $t$, and the related large deviation function (LDF) $\pi(q)$ defined by
\begin{equation} \label{deflargdev}
  \pi(q) = \lim_{t \to \infty} \frac{1}{t} \, \ln P(Q = q \, t,t) \ ,
\end{equation}
and expressed in terms of the fluctuating time-averaged current $q$. 
For finite-size systems $\pi(q)$ is peaked around its maximum
$q_{\text{st}} = \lim_{t \to \infty} \langle Q \rangle / t$, which becomes
sharper as the system size increases (i.e., fluctuations decrease with system 
size). 
In most physical systems, $\pi(q)$ does not edgewise reduce to a parabolic form
--- in other words it describes current fluctuations far beyond a Gaussian 
distribution, even in equilibrium. 
As such, it appears as a promising tool to probe the distinctive features of 
NESS as compared to equilibrium states.

For systems driven out of equilibrium by a field $E$, the fluctuation relation 
takes the generic form $\pi(-q) = \pi(q) - E q$. 
It connects the probabilities of observing longstanding current deviations both
close to and far from the steady average $q_{\text{st}}$.
\medskip

The fluctuations of the longitudinal current are also fully encoded in the 
moments or the cumulants of $Q$. 
We first introduce the generating function
\begin{equation}\label{defZ}
  Z(s,t) = \left\langle \ee^{-s \, Q(t)} \right\rangle =
  \left\langle \exp \left[- s \int_0^t \dd t' \int \dd^dx \, 
  j_\parallel({\bf x},t') \right] \right\rangle \ ;
\end{equation}
moments of $Q$ are then given by the derivatives of $Z(s,t)$ at $s=0$:
\begin{equation}
  (-1)^n \, \frac{\dd^n Z(s,t)}{\dd s^n} \Big|_{s=0} = 
  \left\langle Q(t)^n \right\rangle \ .
\end{equation} 
After resorting to the standard Janssen-De Dominicis mapping to a field theory
\cite{janssen1, dedominicis, bauschjanssenwagner, janssen2}, we arrive for the 
systems presented above at the generic form
\begin{equation} \label{defactionS}
  Z(s,t) = \int{\mathcal D}\phi \, {\mathcal D}\bar{\phi} \ 
  \ee^{-S[\bar{\phi},\phi,s;t]} \ .
\end{equation}
The specific expressions for the actions $S[\bar{\phi},\phi,s;t]$ will
be our starting points in the analysis of the systems presented above.
We utilise in our work the action functional field theory developed in
\cite{janssen1, dedominicis, bauschjanssenwagner}, rather
than the operator formalism of \cite{martinsiggiarose}. These related but
distinct methods lead to quite different techniques, as for instance
recently exemplified in \cite{andersen}.

A few remarks are in order before we proceed with evaluating the large 
deviation function for specific systems. 
Note that the quantity $Z(s,t)$ defined in \eref{defZ}, \eref{defactionS} does
play the role of a partition function, whereas the usual dynamic `partition 
function' $Z(s=0,t) \equiv 1$ (which expresses the conservation of probability)
and carries no relevant information. 
Moreover, since the action defined in \eref{defactionS} does not describe a 
Markov process anymore, it is expected that $Z(s,t)$ will grow exponentially 
with time. 
It is therefore conventional to introduce a time intensive `dynamical free 
energy'
\begin{equation} \label{defmu}
  \mu(s) = \lim_{t \to \infty} \frac{1}{t} \ln Z(s,t) \ ,
\end{equation} 
which is related to the current large deviation function $\pi(q)$ as defined in
\eref{deflargdev} through a Legendre transform,
\begin{equation} \label{eq:legendre}
  \pi(q) = \text{max}_s \{ \mu(s) + s q \} \ .
\end{equation}
Formally, the dynamical free energy $\mu(s)$ is nothing but the generating
function for the cumulants of the integrated longitudinal current $Q$ in the
asymptotic long-time limit:
\begin{equation}
  (-1)^n \, \frac{\dd^n \mu(s)}{\dd s^n} \Big|_{s=0} = 
  \lim_{t \to \infty} \frac{\left\langle Q(t)^n \right\rangle_c}{t} \ .
\end{equation}

\subsection{Steady states with non-zero value of $s$}
\label{subsec:s-states}

In analogy with standard thermodynamics, we observe that the quantity $Z(s,t)$ 
introduced in (\ref{defZ}) facilitates the description of current fluctuations 
in a {\em canonical} way (i.e., at fixed intensive parameter $s$) rather than 
{\em microcanonically} (at fixed value of the time-extensive quantity $Q$).

Besides constituting a useful computational trick, introducing $Z(s,t)$ also
provides us with a physical picture of the very configurations that give birth 
to large deviations. 
Although at $s \neq 0$ the action defined in \eref{defactionS} does not 
correspond to a stochastic process anymore, one can understand 
$S[\bar\phi,\phi,s;t]$ as describing not a single, but rather an {\em ensemble}
of systems evolving in parallel, with dynamical rules that favour histories 
with non-zero longstanding deviation of the particle current
\cite{giardinakurchanpeliti}.  
The value of physical observables such as correlation functions inferred from
$S[\bar\phi,\phi,s;t]$ at $s\neq 0$ can thus be understood as the typical value
of these observables in a {\em modified} steady state wherein the mean current 
is enforced to take an average value $q(s)$ that is different from the average
$q_{\text{st}}$ taken in the steady state.
We see for instance from \eref{defZ} that negative values of $s$ favour 
histories with an excess current in the direction of the field.  
Quantitatively, the correspondence between $q(s)$ and $s$ is the same as in the
Legendre transform \eref{eq:legendre}.
The reader is referred to \cite{giardinakurchanpeliti,thermo-formalism} for 
further examples and for a numerical exploitation of this picture.

\section{Driven diffusive system with mutually exclusive particles and the 
 	 noisy Burgers equation}

\subsection{Field-theoretic formulation}

We wish to determine the distribution function of the integrated
longitudinal current. 
As outlined in section~\ref{subsec:ldf}, this amounts to determine the
`dynamical free-energy' $Z(s,t)$ defined in \eref{defZ}, rewritten as 
$Z(s,t) = \int{\mathcal D} \phi{\mathcal D} \, \bar{\phi} \ 
\ee^{-S[\bar{\phi},\phi,s;t]}$. 
Using periodic boundary conditions we find that the dynamical action 
$S[\bar\phi,\phi,s;t]$ is given by
\begin{eqnarray}\label{actionS}
  \fl \qquad S[\bar\phi,\phi,s;t] = &\int\dd^d x \int_0^t \dd t' \, \bigg[ 
  \bar{\phi} \left( \p_{t'} - D \nabla_\perp^2 - D \lambda \nabla_\parallel^2 
  \right) \phi - D \left( \nabla_\perp \bar{\phi} \right)^2 
  - D \sigma \left( \nabla_\parallel \bar{\phi} \right)^2 \nonumber \\ 
  & \qquad\qquad\quad - \frac{D s g}{2} \, \phi^2 
  + \frac{D g}{2} \, (\nabla_\parallel \bar{\phi}) \, \phi^2 \bigg] 
  + s \, j_0 \, L^d t - s^2 \, D \sigma \, L^d t \ .
\end{eqnarray}
(We use the It\^o convention as regards time discretisation.)
It mainly differs from the stochastic action $S[\bar\phi,\phi,s=0;t]$ through a
new quadratic term in the field $\phi$, aside from the two deterministic 
contributions $\sim L^d t$.
\medskip

In the case of non-interacting particles, each of the cumulants 
$\langle Q^n \rangle_c$ is extensive in the system volume, which means that
the corresponding free energy $\mu_{\text{\tiny{indep}}}(s)$ takes the scaling 
form
\begin{equation}
  \mu_{\text{\tiny{indep}}}(s,L) = L^d \, \hat\mu_{\text{\tiny{indep}}}(s) 
  = L^d \, \left( B_1 s + B_2 s^2 + \ldots \right) \ ,
\end{equation}
where $\hat\mu_{\text{\tiny{indep}}}(s)$ has a well-defined power expansion 
around $s=0$.  
It is known from specific examples, in \cite{kim, lebowitzspohn} or out of
equilibrium \cite{derridalebowitz, derridaappert}, that interactions can lead 
to an expansion of $\hat\mu(s) \equiv L^{-d} \mu(s,L)$ involving non-integer 
powers of $|s|$ (in the infinite volume limit). 
Such non-analytic behaviour at $s=0$ simply mirrors the existence of an 
infinite cumulant of the current as $L \to \infty$. 
For instance, the LDF for the totally asymmetric exclusion process (TASEP) in 
the large system limit (with finite density $\rho$) reads 
\cite{derridalebowitz, derridaappert}
\begin{equation} \label{eq:ldfTASEP}
  \hat\mu_{\text{\tiny{TASEP}}}(s) = - \rho (1 - \rho) s 
   + \frac{(3\pi)^{\frac 23}}{5} \, [\rho(1-\rho)]^{\frac 43} |s|^{\frac 53}
   + \ldots \quad \text{for } s < 0 \ ,
\end{equation}
which means that the variance of the particle current grows faster than the 
system size, whereas in the symmetric exclusion process (SEP) one has 
\cite{kim, lebowitzspohn}
\begin{equation} \label{eq:ldfSEP}
  \hat\mu_{\text{\tiny{SEP}}}(s) = \frac 12 \rho (1 - \rho) s^{2} 
  + \frac{2^{\frac 13}(2\pi)^{\frac 23}}{5} \, [\rho(1-\rho)]^{\frac 43} 
  |s|^{\frac 83} + \ldots \quad \text{for } s < 0 \ .
\end{equation}
The anomalous scaling of the current LDF occurs here for a higher-order 
cumulant, reflecting that the anomalous distribution of the current arises from
sharper details than in the TASEP.  
From a general point a view, we expect the exponents in \eref{eq:ldfTASEP} or
\eref{eq:ldfSEP} to be universal (for other models within the same universality
class), while the prefactors should be model-dependent.

In our case, the continuum and large system size limits will be tackled in 
section \ref{subsec:ddsren} with the presentation of the -- unavoidably 
technical -- RG analysis of the driven diffusive system we are considering. 
We then proceed to compute the dynamical free energy $\mu(s)$ in 
subsection~\ref{subsec:eval_mu_dds}, before extracting its asymptotic behavior 
as a function of $s$ in \ref{subsec:scaling_mu_dds}.

\subsection{Renormalisation}
\label{subsec:ddsren}

The presence of the additional quadratic vertex in~$S[\bar\phi,\phi,s;t]$, in 
spite of allowing for new one-loop diagrams contributing to the three-point 
vertex function $\Gamma^{(1,2)}$, does not alter the remarkable property of 
this model that there are strictly no singular loop corrections to 
$\Gamma^{(1,2)}$, i.e., all new contributions lead to singularities that cancel
out precisely. 
One can explicitly check that order by order in the perturbation expansion, or
infer this property from Galilean invariance, see \eref{eq:inv_galilean} below.
The same statements hold for $\Gamma^{(0,2)}$, which receives no singular loop 
correction either.  
However, both $\Gamma^{(1,1)}$ and $\Gamma^{(2,0)}$ have vertex corrections
given by exactly the same diagrams as with $s=0$, but with values that are now 
$s$-dependent. 
As a consequence, no new renormalisation is required as compared with the 
stochastic case $s=0$.  
Note that for $\sigma = \lambda$ the Einstein relation also holds in the
longitudinal sector.
Then the symmetry
\begin{equation}
  S[\bar{\phi}(t),\phi(t)] = S[\bar{\phi}(-t)-\phi(-t),-\phi(-t)] \ ,
\end{equation}
which for $s=0$ arises from the existence of a free energy functional wherefrom
the Langevin equation derived, continues to hold for $s \not= 0$ -- in spite of
the action not representing a stochastic process anymore -- with the 
consequence that the vertex functions $\Gamma^{(1,2)}$ and $\Gamma^{(0,2)}$ 
are given by their tree-level expressions. 

In the remainder of this subsection, we recall for later use the main lines of 
the renormalisation analysis, which is the same as in the original approach
\cite{janssenschmittmann,taeuber}.
Denoting by $\kappa$ an arbitrary momentum scale, the scaling dimensions of the
various fields and coupling constants are fixed by the requirement and 
convenient choice that the action $S$ as well as the diffusivities $D$ and 
$D \lambda$ be dimensionless,
\begin{eqnarray}
  &[x]=\kappa^{-1} \ , \quad [t]=\kappa^{-2} \ , \quad & [q]=\kappa \ , \quad 
  [\omega]=\kappa^2 \ , \quad [\phi]=[\bar\phi]=\kappa^{d/2} \ ,  \\
  &{[D]}=[\sigma]=[\lambda]=\kappa^0 \ , \quad & [g]=\kappa^{1-d/2} \ , 
  \quad {[s]}=\kappa^{1+d/2} \ . 
\end{eqnarray}
Of course, any other choice for the scaling dimensions of the
field $\bar\phi$ and the diffusion constant $D$ would eventually lead to the
same physical consequences. We infer from the scaling dimension of the nonlinearity the upper 
critical dimension $d_c=2$, and also $[s \, g \, D]=\kappa^2$.
Next we define renormalised fields and parameters via
\begin{eqnarray}
  &\phi_{\sR} = Z_\phi^{1/2} \, \phi \ , \quad 
  &{\bar\phi}_{\sR} = Z_{\bar\phi}^{1/2} \, {\bar\phi} \ , \\
  &D_{\sR} = Z_D \, D \ , \quad &\sigma_{\sR} = Z_\sigma \, \sigma \ , \quad
  \lambda_{\sR} = Z_\lambda \, \lambda \ , \quad 
  g_{\sR} = Z_g \, g \, \kappa^{-1+d/2} \ . 
\end{eqnarray}

In the transverse sector, i.e. on the $q_\parallel=0$ momentum shell, the 
action is merely Gaussian and we have to all orders in the perturbation 
expansion
\begin{eqnarray}
  \Gamma^{(1,1)}(\bq_\perp,q_\parallel=0,\omega) &= 
  i \omega + D \, \bq_\perp^2 \ , \\ 
  \Gamma^{(2,0)}(\bq_\perp,q_\parallel=0,\omega) &= -2 D \, \bq_\perp^2 \ .
\end{eqnarray}
As a consequence, $Z_{\bar\phi}^{1/2} \, Z_{\phi}^{1/2} = 1$ and 
$Z_{\bar\phi}=1$. 
Then, exploiting that the transverse sector is in equilibrium, we obtain from
the associated fluctuation--dissipation theorem the relation 
$Z_D = Z_{\bar\phi}^{-1/2} \, Z_{\phi}^{1/2}$.
Thus we see that neither the fields nor $D$ renormalise:
\begin{equation} \label{ddszid}
  Z_{\bar\phi} = Z_{\phi} = Z_D = 1 \ ,
\end{equation}
and consequently infer the {\em exact} transverse Gaussian scaling exponents
\begin{equation}
  \eta = 0 \ , \quad z_\perp = 2 \ .
\end{equation}

The invariance of the action under Galilean transformations
\begin{equation} \label{eq:inv_galilean}
  \phi'(\bx_\perp',x_\parallel',t') = 
  \phi(\bx_\perp,x_\parallel-D g u\, t,t) - u
\end{equation}
shows that $\phi$ and $u$ must transform in the same way under renormalisation
($u$ is the constant velocity of the Galilean transformation).
Moreover, the product $Dgu$ also remains invariant under RG. Thus, 
\begin{equation} \label{galinz}
  Z_g = Z_D^{-1} \, Z_\phi^{-1/2} = 1 \ ,
\end{equation}
and the coupling constant $g$ is not renormalised either. 

\begin{figure}
\begin{center}
\includegraphics[width=0.5 \textwidth]{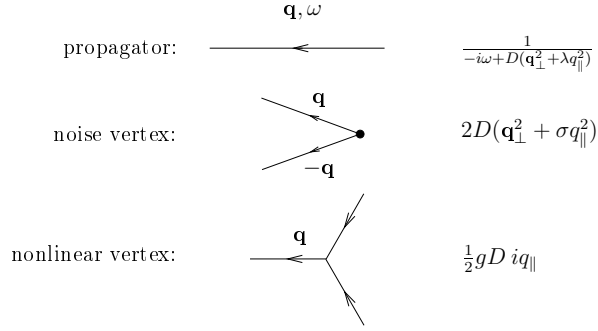}
\caption{Propagator, noise vertex, and nonlinear vertex used in the 
   perturbation expansion for the driven lattice gas.}
\label{fig:verticespropagator}
\end{center}
\end{figure}
We now have to perform an explicit loop expansion using the propagator and 
vertices depicted in Fig.~\ref{fig:verticespropagator}. 
We use the dimensional regularisation scheme and perform the expansion in 
powers of $\varepsilon = d_c - d = 2 - d$.
A straightforward explicit computation of the two-point vertex functions 
yields
\begin{eqnarray}
  &Z_\lambda &= 1 + \frac{v_{\sR}}{16 \varepsilon} \, (3+w_{\sR}) 
  + \mathcal O (v_{\sR}^2) \ , \\
  &Z_\sigma &= 1 + \frac{v_{\sR}}{32 \varepsilon} 
  \left( 3w_{\sR}^{-1}+2+3w_{\sR} \right) + \mathcal O (v_{\sR}^2) \ ,
\end{eqnarray}
where we have defined renormalised effective couplings
\begin{equation}
  w_{\sR} = \frac{\sigma_{\sR}}{\lambda_{\sR}} \ , \quad 
  v_{\sR} = Z_\lambda^{3/2} \, \frac{g^2}{\lambda^{3/2}} \ C_d \, \kappa^{d-2}
  \ , \quad \text{with } C_d = \frac{\Gamma(2-d/2)}{2^{d-1} \pi^{d/2}} \ ,
\end{equation}
and Wilson's flow functions
\begin{eqnarray}
  &\gamma_\lambda &= \kappa \partial_\kappa \vert_0 \ 
  \ln \frac{\lambda_{\sR}}{\lambda} = - \frac{v_{\sR}}{16} 
  \left( 3+w_{\sR} \right) \ , \\
  &\gamma_\sigma &= \kappa \partial_\kappa \vert_0 \ 
  \ln \frac{\sigma_{\sR}}{\sigma} = - \frac{v_{\sR}}{32} 
  \left( 3w_{\sR}^{-1}+2+3w_{\sR} \right) 
\end{eqnarray}
to first order in $v_{\sR}$ (the subscript `0' here indicates that the 
derivatives are to be performed in the unrenormalised theory).
The RG beta functions follow
\begin{eqnarray}
  \beta_w &= \kappa \partial_\kappa \vert_0 \, w_{\sR} &= w_{\sR} 
  \left( \gamma_\sigma - \gamma_\lambda \right) = - \frac{v_{\sR}}{32} 
  \left( w_{\sR}-1 \right) \left( w_{\sR}-3 \right) \ , \\ 
  \beta_v &= \kappa \partial_\kappa \vert_0 \, v_{\sR} &= 
  v_{\sR} \left( d-2 - \frac 3 2 \, \gamma_\lambda \right) \ . 
  \label{eq:rgbetav}
\end{eqnarray}
RG fixed points are now determined by the zeros of these beta functions.
At any non-trivial fixed point $0<v^\star<\infty$, $w^\star=1$ is stable and 
the system reaches effective equilibrium ($\sigma^\star=\lambda^\star$). 
To {\em all orders} in perturbation theory we then obtain from 
\eref{eq:rgbetav}
\begin{equation}
  \gamma_\lambda^\star = \gamma_\sigma^\star =\frac{2}{3} \, (d-2) \ ,
\end{equation}
in dimensions $d \leq d_c = 2$.

The dynamical correlation function scales as
\begin{equation}
  C(\bq,\omega) = \bq_\perp^{-z_{\perp}-2+\eta} \,
  \hat C\left(\frac{q_\parallel}{|\bq_\perp|^{1+\Delta}},
  \frac{\omega}{|\bq_\perp|^{z}}\right) \ ,
\end{equation}
where in $d \leq 2$ dimensions {\em exactly}
\begin{equation}
  \Delta = -\frac{\gamma^\star_\lambda}{2} = \frac{2-d}{3} \ .
\end{equation}
We thus obtain
\begin{equation}
  z_\parallel = \frac{z_{\perp}}{1+\Delta}=\frac{6}{5-d} \ ,
\end{equation}
and as expected, $z_{(\parallel)} = 3/2$ in one dimension.
For $d > 2$ the system scales towards the Gaussian fixed point $v_0^* = 0$,
whence one arrives at the mean-field scaling exponents $\Delta = 0$, 
$z_\parallel = z_\perp = 2$.

\subsection{Evaluation of the cumulant generating function $\mu(s)$}
\label{subsec:eval_mu_dds}

Having recalled the main results of the RG analysis, we can now turn to the 
evaluation of the dynamical free energy $\mu(s)$.
The dynamical action \eref{actionS} contains an extensive (in time as well) 
deterministic contribution, $L^d t \left(s j_0 - s^2 D \sigma \right)$, and a 
fluctuating one, for which it will prove sufficient to extract the tree level
approximation.
Quite unfamiliarly in dynamical field theory, the latter contribution arises
from the normalisation of the path integral. 
The Gaussian contribution to the action $S[\bar{\phi},\phi,s;t]$ takes the form
$\int \frac 1 2 \ \big(\bar \phi \ \phi \big) \ \Gamma_0 \pmatrix{\bar \phi \cr
\phi}$ with, conveniently written in Fourier space,
\begin{equation}
  \Gamma_0(\bq,\omega) = \left( \begin{array}{cc}
  -2D \left( \bq_\perp^2 + \sigma q_\parallel^2 \right) & 
  -i \omega + D \left( \bq_\perp^2 +\lambda q_\parallel^2 \right) \\
  i \omega + D \left( \bq_\perp^2 + \lambda q_\parallel^2 \right) & -D g s 
  \end{array} \right) \ .
\end{equation}
We have
\begin{eqnarray}
  \det \Gamma_0  & = -\left[ \omega^2 + \Omega_s^2(\bq) \right] \ , \\
  \Omega_s^2(\bq)& = D^2 \left( \bq_\perp^2 + \lambda q_\parallel^2 \right)^2
  - 2 s g D^2 \left( \bq_\perp^2 + \sigma q_\parallel^2 \right) \ ,
\label{eq:freq_abbr}
\end{eqnarray}
and integrating over the fields yields the following tree level contribution to
$\mu(s)$:
\begin{eqnarray}
  &&\fl \int{\mathcal D}\phi \, {\mathcal D}\bar{\phi} \ \exp\left[ - \int \,
  \frac 1 2 \ \big(\bar\phi \ \phi \big) \ \Gamma_0 \pmatrix{\bar\phi \cr \phi}
  \right] \\
  &&\fl \qquad = \exp \left[ - \frac 1 2 L^d t \int \frac{\dd^d q}{(2\pi)^d} \,
  \frac{\dd \omega}{2\pi} \ \ln \left| \det 
  \frac{\Gamma_0(\bq,\omega)}{\Gamma_0(\bq,\omega)|_{s=0}} \right| \right] \\
  &&\fl \qquad = \exp \left[ - \frac 1 2 L^d t \int \frac{\dd^d q}{(2\pi)^d} \,
  \frac{\dd \omega}{2\pi} \ln \left( 1 - \frac{2 s g D^2 (\bq_\perp^2 + 
  \sigma q_\parallel^2)}{\omega^2+D^2 (\bq_\perp^2 + \lambda q_\parallel^2)^2}
  \right) \right] \\
  &&\fl \quad = \exp\left( - \frac 1 2 L^d t D \! \int \frac{\dd^d q}{(2\pi)^d}
  \left[ \sqrt{\left( \bq_\perp^2 + \lambda q_\parallel^2 \right)^2 
  - 2 s g \left( \bq_\perp^2 + \sigma q_\parallel^2 \right)} 
  - \left( \bq_\perp^2 + \lambda q_\parallel^2 \right) \right] \right) .
  \label{eq:integ_mu}
\end{eqnarray}

At the RG fixed point the values of $\lambda$ and $\sigma$ are not fixed, but 
become equal ($w^{\star}=1$). 
Without loss of generality, one can thus start from equal bare coupling 
constants $\sigma = \lambda$ (as was for instance implicitly assumed in 
\cite{vanbeijerenkutnerspohn}).
The last integral (\ref{eq:integ_mu}) can now be carried out explicitly, and 
its infrared divergence is isolated by extracting the first term of the small 
$s$ expansion.
This gives
\begin{equation} \label{eq:mu_of_s_with_integ}
  \fl \qquad 
  L^{-d} \, \mu(s) = -j_0 s + s^2 D \sigma + \frac 1 2 s g D \Lambda^d 
  + \frac{D \lambda^{-1/2} (-gs)^{1+d/2} \, \Gamma(d) \, \Gamma(-1-d/2)}
  {2^d (2\pi)^{d/2} \, \big[\Gamma(d/2)\big]^2} \ .
\end{equation}

\subsection{Scaling behaviour of $\mu(s)$}
\label{subsec:scaling_mu_dds}

To determine the expansion of $\mu(s)$ in powers of $s$ in the continuum limit,
and at the stable RG fixed point $w^\star = 1$ ($\lambda = \sigma$) we notice 
that making explicit the $1/\varepsilon$ pole in \eref{eq:mu_of_s_with_integ},
\begin{equation} 
  L^{-d}\, \mu(s) = - j_0 s + s^2 D \lambda \left[ 1 + \frac{C_d g^2}
  {\lambda^{3/2}} \, \frac{(-g s)^{-\varepsilon/2}}{4 \varepsilon} \right] \ ,
\end{equation}
one recognises the expression for $Z_\lambda = 1+ \frac{v_{\sR}}{4\varepsilon}$
and finds (at the convenient normalisation point $\kappa = (-sg)^{\frac 1 2}$)
\begin{equation} 
  L^{-d} \, \mu(s) = - j_0 s +  D \lambda_{\sR} s^2 \ .
\end{equation}
Finally, using that in the vicinity of the RG fixed point
$\lambda_{\sR} \sim \lambda \kappa^{\gamma_\lambda^\star}$, with the anomalous
scaling dimension $\gamma_\lambda^\star = \gamma_\sigma^\star = 
- \frac23 \varepsilon$ of the noise correlator, the asymptotic scaling of 
$\mu(s)$ as a function of $s$ reads (below the critical dimension $d_c = 2$)
\begin{equation} \label{eq:result_mu_s}
  L^{-d} \, \mu(s) = - j_0 s + {\mathcal A}_d \, s^{2 - \varepsilon/3} 
  = - j_0 s + {\mathcal A}_d \, s^{(d + 4) / 3} \ .
\end{equation}
The constant prefactor ${\mathcal A}_d$ is non-universal, contrary to the 
exponent $2 - \varepsilon / 3 = (d + 4) / 3$, which holds exactly to all orders
in perturbation theory.  
In one dimension, we recover the exponent $5/3$ of the totally asymmetric 
exclusion process \cite{derridalebowitz,derridaappert}, see \eref{eq:ldfTASEP}.
This results expectedly corroborates that the TASEP and systems described by 
the Burgers equation belong to the same (dynamical) universality class.

We emphasize that the scaling exponent in \eref{eq:result_mu_s} merely follows 
from the noise renormalisation and could thus have been obtained through 
straightforward dynamic scaling. 
A simpler, though less explicit, way to arrive at that conclusion is to employ 
the matching condition stemming from the $\Lambda^d$ IR divergence in 
\eref{eq:mu_of_s_with_integ}:
The only factor acquiring an anomalous dimension in the crucial second term in
\eref{eq:mu_of_s_with_integ} is the parameter $\sigma$; with 
$\gamma_\sigma^\star = \frac{2}{3} \, (d-2)$ and the matching condition 
$\kappa \sim (- s g)^{1/2}$ that directly follows from the scaling dimensions,
we again arrive at $L^{-d} \, \mu(s) + j_0 s \sim s^{2 + (d-2)/3}$.
Indeed, this will be our method of choice in the following subsection as well 
as for the remaining two systems.

\subsection{Logarithmic corrections to $\mu(s)$ in two dimensions}

Precisely at the upper critical dimension $d_c=2$, the RG flow equations 
provide access to the logarithmic corrections to the mean-field critical power 
laws. 
In our context, they also specify the logarithmic correction to the $s^2$ term 
in $\mu(s)$.  
To determine how the renormalised coupling constants vary under scale 
transformation $\kappa\mapsto\ell\kappa$, let us recall that running couplings 
$\tilde{\lambda}(\ell)$ and $\tilde{v}(\ell)$ are defined through
\begin{equation}
  \ell \frac{\dd\tilde\lambda(\ell)}{\dd\ell} = \gamma_\lambda(\ell) \,
  \tilde\lambda(\ell) \ , \quad 
  \ell \frac{\dd\tilde v(\ell)}{\dd\ell} = \beta_v(\ell) \ ,
\end{equation}
with initial conditions $\tilde{\lambda}(1)=\lambda_{\sR}$, 
$\tilde{v}(1)=v_{\sR}$, and where 
$\gamma_\lambda(\ell) = \gamma_\lambda(\tilde v(\ell))$,
$\beta_v(\ell) = \beta_v(\tilde v(\ell))$.
In $d_c=2$ dimensions, the flow equation for $\tilde v(\ell)$ reads
\begin{equation}
   \ell \frac{\dd\tilde v(\ell)}{\dd\ell} = \frac 38 \, \tilde v^2(\ell) 
   + \mathcal O (\tilde v^3(\ell)) \ ,
\end{equation}
which is solved by
\begin{equation}
   \tilde v(\ell) = \frac{v_{\sR}}{1 - \frac 38 v_{\sR} \ln \ell} \ .
\end{equation}
Upon inserting this result in the flow equation for $\tilde{\lambda}(\ell)$, 
one obtains
\begin{equation}
   \tilde\lambda(\ell) \sim \bar\lambda \left(\ln \ell\right)^{2/3} \ ,
\end{equation}
which gives the precise form of \eref{eq:result_mu_s} at $d_c=2$:
\begin{equation}
   L^{-2} \, \mu(s) = 
   - j_0 s + {\mathcal A}_2 \, s^2 \big(-\ln |s| \big)^{2/3} \ .
\end{equation}
Of course, such logarithmic corrections are familiar within the framework of
RG calculations at the upper critical dimension.
Yet we stress that this result constitutes the first {\em exact} result for the
LDF scaling of a two-dimensional system, where the methods of integrable 
systems fail to apply.

\subsection{Correlation functions}
\label{subseq:correlation_DDS}

\begin{figure}
\begin{center}
\includegraphics[width=0.4 \textwidth]{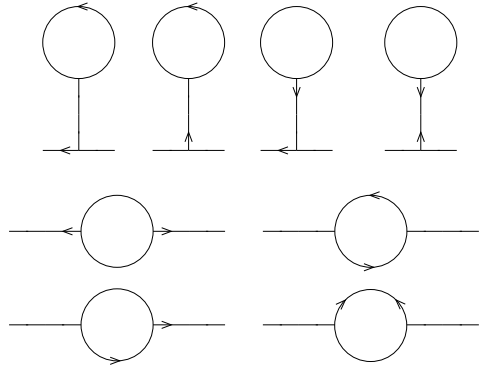}
\caption{One-loop graphs which renormalise the correlation function in the 
   $s$-state.}
\label{fig:graph_corridor}
\end{center}
\end{figure}
\begin{figure}
\begin{center}
\includegraphics[width=0.25 \textwidth]{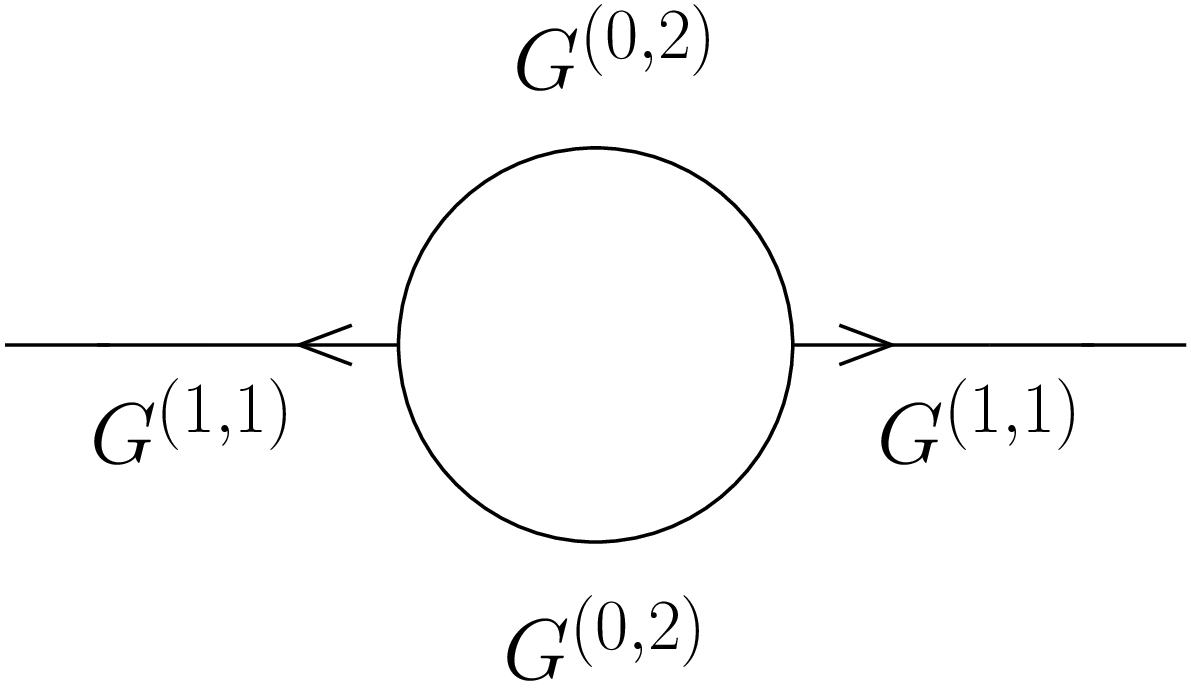}
\caption{Single diagram contributing to $C(\bq,t,t)$ in the $s$-state, at 
   minimal order with respect to the external momentum $\bq$. The expressions 
   for the propagators are given in \eref{eq:propagators_s-state}.} 
\label{fig:contributing_loop}
\end{center}
\end{figure}

In this subsection, we aim at determining how the correlation function are 
modified for small (negative) values of $s$. 
As announced in section~\ref{subsec:s-states}, states with $s<0$ are 
characteristic of histories with excess current in the direction of the driving
field. 
Let $G^{(m,n)}=\langle\bar{\phi}^m\phi^n\rangle$ denote the m-$\bar{\phi}$ and 
n-$\phi$ correlation function. 
For the complete quadratic part of the action, we find that
\begin{eqnarray}
  G^{(1,1)}(\bq,t,t') = \overbrace{\Theta(t'-t) \, \frac{\Omega_0+\Omega_s}
  {2\Omega_s} \, \ee^{-\Omega_s(t'-t)}}^{\text{causal part}} + 
  \overbrace{\Theta(t-t') \, \frac{\Omega_0-\Omega_s}{2\Omega_s} \,
  \ee^{-\Omega_s(t-t')}}^{\text{non causal part}} \, , \nonumber \\
  G^{(0,2)}(\bq,t,t') = \frac{D(\bq_\perp^2+\sigma q_\parallel^2)}{\Omega_s} \,
  \ee^{-\Omega_s|t'-t|} \ , \label{eq:propagators_s-state} \\
  G^{(2,0)}(\bq,t,t') = \frac{D s g}{2\Omega_s} \, \ee^{-\Omega_s|t'-t|} \ ,
  \nonumber
\end{eqnarray}
with the abbreviation \eref{eq:freq_abbr}, and with the convention that the 
step function satisfies $\Theta(0)=0$. 
(Note that the above calculations hold only for $s<0$). 
The diagrammatic expansion contains no causality constraint on the loops 
anymore. 
The long-range spatial correlations in the $s$-state are obtained from an 
expansion in $\bq$ at minimal order. 
To one loop, the correlation function is renormalised through the graphs shown 
in Fig.~\ref{fig:graph_corridor}. 
To lowest order in $\bq$, we find that all contributions cancel except for the 
single loop diagram depicted in Fig.~\ref{fig:contributing_loop}, which yields
\begin{eqnarray}
  &C(\bq,t,t) - C(\bq,t,t) \big|_{s=0} &= (-2gs)^{-1/2} \left( \bq_\perp^2 + 
  \lambda q_\parallel^2 \right)^{1/2} \nonumber \\
  &&\times \left[ 1 + g^2 \lambda^{-1/2} \, (-gs)^{-\varepsilon/2} \,
  \frac{1}{16\pi\varepsilon} \, \frac{q_\parallel^2}
  {\bq_\perp^2+\lambda q_\parallel^2} \right] \, .
\end{eqnarray}
Averaging along the equilibrium directions, i.e., at $q_{\perp}=0$, one 
recognises
\begin{equation}
  C(q_{\parallel},0,t,t) - C(q_{\parallel},0,t,t) \big|_{s=0} = 
  (-2gs)^{-1/2} \, \lambda_{\sR}^{1/2} \, q_{\parallel} \ ,
\end{equation}
and finally obtains the exponent of $s$ in the $s$-correlation function:
\begin{equation}
  C(q_{\parallel},0,t,t) - C(q_{\parallel},0,t,t) \big|_{s=0} \propto 
  (-s)^{-1/2 - \varepsilon/6} \ .
\end{equation}
In one dimension, this gives the power law $(-s)^{-2/3}$, which, to our 
knowledge, has not been obtained through exact approaches for systems in the 
Burgers/KPZ universality class.

\section{A driven diffusive system with a continuous phase transition}

\subsection{Renormalisation}

The previous considerations and calculations can readily be generalised to the
critical driven diffusive system described by the nonlinear Langevin equation
\eref{drblan} with noise correlator \eref{ddsnoi}.
Since only the transverse sector in momentum space is rendered critical, we
merely need to replace the propagator in Fig.~\ref{fig:verticespropagator}
with the expression $1 / [-i \omega + D \bq_\perp^2 (r + \bq_\perp^2) 
+ D \lambda \, q_\parallel^2 ]$.
Dimensional analysis yields $[r] = \kappa^2$, and at criticality ($r = 0$)
$[q_\parallel] = [q_\perp]^2 = \kappa^2$, and hence $[\omega] = \kappa^4$.
This yields for the scaling dimension of the nonlinear coupling 
$[g] = \kappa^{(5-d)/2}$, wherefrom we infer the upper critical dimension
$d_c = 5$.

Renormalising this model proceeds in much the same way as described above in
section~\ref{subsec:ddsren} (see \cite{schmittmannzia1,taeuber}). 
Since again the nonlinear vertex is proportional to
$i q_\parallel$, the transverse sector is Gaussian with
\begin{eqnarray}
 \Gamma^{(1,1)} (\bq_\perp,q_\parallel=0,\omega) &= i \omega + D \bq_\perp^2 
	(r + \bq_\perp^2) \ , \\
 \Gamma^{(2,0)} (\bq_\perp,q_\parallel=0,\omega) &= -2 D \bq_\perp^2 \ .
\end{eqnarray}
Consequently \eref{ddszid} holds here as well, implying that
\begin{equation}
  \eta = 0 \ , \quad \nu = \frac 12 \ , \quad z_\perp = 4
\end{equation}
{\em exactly}.
The system is of course still invariant with respect to Galilean 
transformations, whence \eref{galinz} is valid. The RG beta function for the 
effective nonlinearity $v = g^2 / \lambda^{3/2}$ now reads
\begin{equation}
  \beta_v = v_{\sR} \left( d-5 - \frac 3 2 \, \gamma_\lambda \right) \ ,
\end{equation}
and thus again to {\em all} orders in the perturbation expansion
\begin{equation}
  \gamma_\lambda^\star = \frac 2 3 \, (d-5) \ ,
\end{equation}
and consequently
\begin{equation}
  \Delta = 1 - \frac{\gamma_\lambda^\star}{2} = \frac{8-d}{3} \ , \quad
  z_\parallel = \frac{z_\perp}{1 + \Delta} = \frac{12}{11-d} \ .
\end{equation}

\subsection{Evaluation and scaling behaviour of $\mu(s)$}

The Gaussian contribution to the action $S[\bar{\phi},\phi,s;t]$ now becomes
\begin{equation}
   \exp \Biggl[ - \frac 1 2 \, L^d t \int \frac{\dd^d q}{(2\pi)^d} \,  
   \frac{\dd \omega}{2\pi} \, \ln \Biggl( 1 - \frac{2 s g D^2 (\bq_\perp^2
   + \sigma q_\parallel^2)}{\omega^2 + D^2 \bigl[ \bq_\perp^2 (r + \bq_\perp^2)
   + \lambda q_\parallel^2 \bigr]^2} \Biggr) \Biggr] .
\end{equation}
Thus we observe that we require the scaling dimension of the noise strength
$\sigma$, which is an {\em irrelevant} parameter, in the RG sense, for this 
model.
A straightforward one-loop calculation gives 
\begin{equation}
  \gamma_\lambda = - v_{\sR} \ ,\quad \gamma_\sigma = 2 - \frac{v_{\sR}}{3} \ ,
\end{equation}
so for $\epsilon = 5 - d > 0$,
\begin{equation}
  \gamma_\sigma^\star = 2 - \frac{2 \epsilon}{9} + \mathcal O (\epsilon^2) \ .
\end{equation}
However time now scales according to $[t] = \kappa^{-4}$, and the appropriate 
matching condition becomes $\kappa \sim (-s g)^{1/4}$.
Thus we finally arrive at
\begin{equation}
    L^{-d} \, \mu(s) = - j_0 s + {\mathcal B}_d \, s^{2 - \epsilon / 18} \ ,
\end{equation}
with a non-universal amplitude ${\mathcal B}_d$.
Note that this result holds only to first order in the dimensional expansion 
for $d = 5 - \epsilon < 5$.
\medskip

At the critical dimension $d_c = 5$, 
\begin{equation}
   \tilde v(\ell) = \frac{v_{\sR}}{1 - \frac 32 v_{\sR} \ln \ell} \ , \quad
   \tilde\sigma(\ell) \sim \bar\sigma \, \ell^2 \left(\ln \ell\right)^{2/9} \ ,
\end{equation}
wherefrom we obtain the logarithmic correction
\begin{equation}
    L^{-5} \, \mu(s) = - j_0 s 
    + {\mathcal B}_5 \, s^2 \big( -\ln |s| \big)^{2/9} \ .
\end{equation}

\section{Superdiffusion in a random velocity field}

\subsection{Mesoscopic formulation}

Before embarking on the determination of $\mu(s)$ for the case of 
(super-)diffusive particles subject to a random velocity field, we have to 
establish how the random field $\psi$ intervenes in the definition of the LDF 
of interest.
The physical features of the current fluctuations are contained in the 
disorder-averaged {\em cumulants} $\overline{\langle Q(t)^n \rangle_c}$.
Here and below the averaging over thermal disorder will be noted by 
$\langle\ldots\rangle$, while the average over the random field will be 
indicated with an overbar, $\overline{\,\vphantom{|}\ldots\,}$.
We are thus ultimately interested in determining the LDF
\begin{equation} \label{eq:mu_of_s_disorder}
  \mu(s) = \lim_{t \to \infty} \frac 1t \, \overline{\,\ln Z_\psi( s,t )\,} \ .
\end{equation}
The disorder-dependent partition function $Z_\psi(s,t)$ reads
\begin{equation} \label{eq:Z_of_s_psi}
  Z_\psi(s,t) = \Big\langle \exp\big[- s \int_0^t \dd t' \, \dd^d x \,
  j_\parallel({\bf x},t') \big] \Big\rangle \ ,
\end{equation}
and as in the previous models, a field-theoretic reformulation of 
\eref{eq:honk_currents}--\eref{eq:honk_noise} and \eref{eq:Z_of_s_psi} enables 
us to rewrite the partition function as
\begin{equation} \label{defZ2}
  Z_\psi(s,t) = \int{\mathcal D}\bar\phi \, {\mathcal D}{\phi} \,
  \ee^{-S_\psi[\bar{\phi},\phi,s;t]} \ ,
\end{equation}
where the quenched and $s$-dependent action is given by
\begin{eqnarray} \label{eq:action_honk}
  \fl \qquad S_\psi[\bar{\phi},\phi,s;t] = & - s^2 D_\parallel L^d t 
  + \int\dd^d x \int_0^t \dd t' \Big[ \bar{\phi} \left( 
  \p_{t'} - D_\perp \nabla_\perp^2 - D_\parallel \nabla_\parallel^2 - \psi \, 
  \nabla_\parallel \right) \phi \nonumber\\  
  &\qquad\qquad\qquad\qquad\qquad\quad - D_\perp (\nabla_\perp \bar{\phi})^2 
  - D_\parallel (\nabla_\parallel \bar{\phi})^2 \, - s \psi \phi \Big] \, .
\end{eqnarray}

The form of this action is related to Honkonen's original ($s=0$) action 
\cite{honkonen} through the canonical transformation
\begin{equation}
  \bar\phi = \bar \rho \ , \quad \phi = \rho + \bar \rho \ ,
\end{equation}
whereupon $S_\psi$ takes the form 
\begin{eqnarray} \label{eq:action_honk2}
  &S_\psi[\bar{\rho},\rho,s;t] = -s^2 D_\parallel L^d t \nonumber \\
  &\quad + \int\dd^d x \int_0^t\dd t' \Bigl[ \bar{\rho} \left( 
  \p_{t'} - D_\perp \nabla_\perp^2 - D_\parallel \nabla_\parallel^2 - \psi \,
  \nabla_\parallel \right) \rho \, + s (\bar\rho + \rho) \psi \Bigr] \, .
\end{eqnarray}
For $s = 0$ this is precisely the action that was studied by Honkonen 
\cite{honkonen}.

In a microscopic lattice gas formulation (independent particles diffusing in a 
random field), $Z_\psi(s,t)$ can also be obtained from the standard Doi-Peliti
approach \cite{pathinteg-stoch}.
Up to irrelevant terms, one arrives again at the action \eref{eq:action_honk},
with the field $\phi$ being related to the original operators through a 
Cole--Hopf transformation.
This illustrates that, either in the Janssen-De Dominicis or in the Doi-Peliti 
manner, the well-tried functional methods of statistical physics can be readily
extended to tackle large deviation functions and the associated $s$-modified 
states, as previously proposed in \cite{vanwijlandracz}.

\subsection{Renormalisation}

The theory has two superficially divergent graphs which lead to the same 
renormalisation of $D_\parallel$. 
As can be seen from direct inspection of possible diagrams, the additional 
$s \psi \phi$ term in the action does not lead to further renormalisation, as
compared to the $s=0$ case. We define as in \cite{honkonen} renormalised 
couplings through $D_\parallel = Z D_{\parallel,{\sR}}$, 
$\lambda = \lambda_{\sR} \kappa^\epsilon$.
The perturbative expansion is organised in powers of a single dimensionless 
parameter $u_{\sR}$ that we fix to 
\begin{equation}
  u_{\sR} = \frac{\lambda_{\sR}}{2 \pi D_{\parallel,{\sR}} D_\perp} \ .
\end{equation}
Examination of the one-loop contribution to $\Gamma^{(1,1)}$ leads to
\begin{equation}
  Z = 1 - \frac{u_{\sR}}{\epsilon} \ , \quad 
  \beta_u = u_{\sR} (u_{\sR}-\epsilon) \ .
\end{equation}
Below the critical dimension $d_c=3$, the RG flow has a single non-trivial
fixed point $u_{\sR}^\star = \epsilon$, at which the system exhibits 
superdiffusion \cite{honkonen}, namely
\begin{equation}
  \langle x_\parallel^2 \rangle \sim t^{1 + \epsilon/2} \ ;
\end{equation}
in other words, the dynamical exponent in the longitudinal direction is 
\begin{equation}
  z_\parallel = \frac{4}{2 + \epsilon} \ .
\end{equation}

\subsection{Determination of $\mu(s)$ }

From a dynamical point of view, the study of disordered systems does not pose 
any particular problem since conservation of probability ensures that the 
dynamical partition function remains equal to $1$ as time evolves. 
This has the well-known consequence that quenched disorder can be treated 
without resorting to the replica trick. 
However, in the present case, our interest aims precisely at the nonzero and 
$s$-dependent dynamical partition function $Z_\psi(s,t)$, which grows 
exponentially with time, with a disorder-dependent rate. 
The dynamical free energy we pursue, denoted by $\mu(s)$, is obtained as the 
average of this rate, see \eref{eq:mu_of_s_disorder}. 
The replica trick represents one convenient means to achieve this disorder 
averaging and to determine $\mu(s)$ via
\begin{equation} \label{eq:mu_of_s_replica}
  \mu(s) = \lim_{t \to \infty}\frac 1t \, \frac{\partial}{\partial n} 
  \Bigg|_{n=0} \!\!\!\! \overline{\, [Z_\psi( s,t )]^n } \ .
\end{equation}

In contrast with previous work \cite{bouchaudetal87} using replicas to probe
dynamical aspects of disordered systems, our approach leads to non-trivial 
couplings between replicas.
Integration over $\psi$ yields
\begin{equation}
  \overline{\, [Z_\psi(s,t)]^n} = \int {\mathcal D}\bar\phi_a \, 
  {\mathcal D}{\phi_a} \, \ee^{-S_n [\bar{\phi},\phi,s;t]} \ ,
\end{equation}
with an effective action $S_n[\bar{\phi_a},\phi_a,s;t]$ whose quadratic part 
reads
\begin{eqnarray}
  \fl \qquad S_n[\bar{\phi_a},\phi_a,s;t] = & -nL^d t D_\parallel s^2 - s^2 
  \frac{\lambda}{2} \sum_{a,b} \int \frac{\dd^{d-1} \bq_\perp}{(2\pi)^{d-1}} \,
  \phi_a(\bq_\perp,0,0) \, \phi_b(-\bq_\perp,0,0) \nonumber \\ 
  & +\sum_a \int \frac{\dd^d \bq}{(2\pi)^d} \, \frac{\dd \omega}{(2\pi)} \Big[
  \bar\phi_a (-i\omega + D \cdot \bq^2) \phi_a - (D \cdot \bq^2) \bar\phi_a^2 
  \Big] \label{eq:S_n_eff} \ ,
  \end{eqnarray}
where we have set $D \cdot \bq^2 = 
D_\perp(\bq_\perp)^2 + D_\parallel (q_\parallel)^2$ for clarity. 
Integration over the noise has given birth to a uniform coupling between 
replicas of the field $\phi$, in the $\omega=0$, $q_\perp=0$ sector. 
However, no new renormalisation is needed, as can be directly checked.
\medskip

The free energy $\mu(s)$ is obtained as the ratio of the determinant of the 
action at $s \neq 0$ and $s=0$.
In our case, the effective action \eref{eq:S_n_eff} is diagonal in Fourier 
space, and the only modes which depend on $s$ are those with $q_\parallel=0$ 
and $\omega=0$. 
We thus require the matrix determinant
\begin{equation} \label{eq:Omega_det}
  \Delta_n(\bq_\perp) = \det \left( \begin{array}{cc}
  -2 D_\perp\bq_\perp^2 \, \mathbb I_n & 
  D_\perp\left(\bq_\perp^2+m^2\right) \, \mathbb I_n \\
  D_\perp \left(\bq_\perp^2+m^2\right) \, \mathbb I_n & 
  -\lambda s^2 \,\mathbb J_n \end{array} \right) \ ,
\end{equation}
where $\mathbb I_n$ denotes the $n \times n$ identity matrix, and $\mathbb J_n$
represents the matrix whose entries are all $1$.
For later convenience, we have also added a mass $m$ to the $\bar\phi \phi$ 
term.
Using standard results about block determinants, we find 
\begin{equation}
  \Delta_n(\bq_\perp) = (-1)^n \left(D_\perp\bq_\perp^2\right)^{2n} \, 
  \Big[ D_\perp^2 \left( \bq_\perp^2+m^2 \right)^2 
  - 2 n\lambda s^2 D_\perp\bq_\perp^2 \Big] \ .
\end{equation}
The corresponding (tree-level) contribution to $\mu(s)$ is then given by
\begin{eqnarray}
  -\frac 12 L^d \, \frac{\partial}{\partial n} \Bigg|_0 & 
  \int \frac{\dd^{d-1} \bq_\perp}{(2\pi)^{d-1}} \, 
  \ln \frac{\Delta_n(\bq_\perp)}{\Delta_n(\bq_\perp)|_{s=0}} = L^d \, 
  \frac{\lambda s^2}{D_\perp} \int \frac{\dd^{d-1}\bq_\perp}{(2\pi)^{d-1}} \,
  \frac{q_\perp^2}{\left( q_\perp^2+m^2 \right)^2} \nonumber \\ 
  &\qquad = L^d \, \frac{\lambda s^2 (2-\epsilon)\, m^{-\epsilon}}
  {4(2\pi)^{2-\epsilon}} \, \Gamma\left( \frac \epsilon 2 \right) 
  \Gamma\left(1-\frac \epsilon 2\right) \ .
\end{eqnarray}
Combining this result with the deterministic contribution to $\mu(s)$ we 
recognise
\begin{equation} \label{eq:mubareHonk}
  \mu(s) = L^d D_\parallel \, s^2 \left( 1 + \frac{u_{\sR}}{\epsilon} \right) 
  = L^d D_{\parallel,{\sR}} \, s^2 \ .
\end{equation}

\subsection{Scaling behaviour of $\mu(s)$}

As in the previous examples, we infer from \eref{eq:mubareHonk} that the 
continuum limit in $\mu(s)$ is determined by the renormalisation of the noise 
coupling constant.
Using $\gamma_{D_\parallel}^\star = - \epsilon$, one gets
\begin{equation}
  L^{-d} \mu(s) \ \sim \ s^2 \kappa^{-\epsilon} \ ,
\end{equation}
but from the explicit expression \eref{eq:Omega_det}, the normalisation point
depends on $s$ only through $\lambda s^2$. 
Since $[\lambda s^2] = \kappa^5$, we obtain the scaling behaviour, valid below 
the critical dimension $d_c = 3$,
\begin{equation} \label{eq:muHonk}
  L^{-d} \mu(s) \ \sim \ s^{2 - 2 \epsilon / 5} \ .
\end{equation}
We note that although this system is in equilibrium, the expansion 
\eref{eq:muHonk} implies that that in a finite-size system the variance of the 
current grows with the linear size $L$ as $L^3$ for $d < 3$, in contrast to
the $L^{d}$ behaviour observed in normally diffusive systems. 
Again, this anomalous behaviour is exemplified here for the first time.
\medskip

At the critical dimension $d_c=3$, the running coupling 
\begin{equation}
  \tilde D_\parallel(\ell) \sim D_{\sR} \ln \ell \ ,
\end{equation}
which gives the logarithmic correction
\begin{equation}
  \mu(s) \sim s^2 (-\ln |s|) \ .
\end{equation}
At the upper critical dimension too, the RG analysis leads to an asymptotically
exact result.

\subsection{Correlation function}
\label{subseq:correlation_honk}

As well as allowing access to free energies, the replica trick can also be used
to determine the correlation functions, averaged over thermal fluctuations and 
the random field. 
This is achieved by noting that
\begin{equation}
  \overline{\langle \phi(\bq_\perp,0,0) \phi(-\bq_\perp,0,0)\rangle_c} = 
  \overline{\langle \phi_a(\bq_\perp,0,0) \phi_a(-\bq_\perp,0,0) \rangle_c 
  \big|_{n \to 0}} \ .
\end{equation}
Thus we obtain at the tree level the correlation function in the $s$-state as
\begin{eqnarray}
  \langle \phi(\bq_\perp,0,0) \phi(-\bq_\perp,0,0)\rangle_c &= 
  \frac{2}{D_\perp \bq_\perp^2} \left( 
  1 - \lambda s^2 \frac{2}{D_\perp \bq_\perp^2} \right) \ .
\end{eqnarray}
The spatio-temporal correlation function reads
\begin{equation} \label{eq:C_treelevel_Honk}
  C(\bx,t)\ =\ C(\bx,t)\big|_{s=0} - \frac{2\lambda s^2}{D_\perp} 
  \int \frac{d^{d-1}}{(2\pi)^{d-1}} \, \frac{1}{\bq_\perp^2} \, 
  \ee^{i\bx_\perp\bq_\perp} \ ,
\end{equation}
i.e., the supplementary correlations in the $s$-steady state are still uniform 
in time and along the longitudinal direction, which means that states conveying
a (slightly) atypical value of the current do not break theses symmetries of
the steady state. 
Including loop corrections to the correlation function does not alter this 
result, but affects the power $s^2$ in \eref{eq:C_treelevel_Honk}, as a
consequence of superdiffusivity.

\section{Conclusions}


In the previous sections, we have shown that the implementation of dynamical RG
techniques to the already well-studied (see refs \cite{evanscohenmorriss} 
through \cite{bertinidesole4}) dynamical free energy $\mu(s)$ associated with 
the particle current provides new insights into the singular dynamical 
\emph{macroscopic limit} of large systems.  
As is the case for the static free energies in thermal equilibrium setups, 
$\mu(s)$ exhibits singularities only in the infinite volume limit. 
In the systems presented in this work, $\mu(s)$ becomes non-analytic at $s=0$, 
where $s$ is the parameter canonically conjugated to the particle current:
\begin{equation}
   L^{-d} \mu(s) + j_0 s \ \sim \ s^a \ , \quad \text{with } \ 1 < a \leq 2 \ .
\end{equation}
In finite-size systems, the variance of the integrated current grows with the 
system size faster than linearly. 
We have shown for the systems under consideration here that the exponents $a$ 
were simply related to the respective dynamical exponents $z$ through the 
anomalous exponent for the noise strengths. 
In our examples, the divergence of the variance ($a < 2$) emerges as a direct 
consequence of superdiffusion ($z < 2$). 
 
Our second system provides an example for a current LDF in presence of twofold,
namely static {\em and} dynamic criticality.
Our final model system constitutes the first example of an {\em equilibrium} 
system for which the fluctuations of the particle current were shown to produce
anomalous scaling with system size.

In previous works, which remained confined to a specific one-dimensional 
system, namely the asymmetric exclusion process, an exponent related to $a$ was
determined using the Bethe ansatz \cite{derridalebowitz, derridaappert,
thermo-formalism}. 
Of course, some of these existing results can be recovered through our 
field-theoretic methods employed here (such as the scaling exponent in $d=1$), 
but we have obtained additional asymptotically exact results in higher 
dimensions. 
As well as opening the path to a wide range of applications, our method also 
connects to other multi-purpose instruments of statistical physics (such as
replicas and supersymmetry for disordered systems) that can be deftly adapted 
to describe the $s$-dependent states. 
As illustrated in sections~\ref{subseq:correlation_DDS} 
and~\ref{subseq:correlation_honk}, our method also gives access to the 
correlation functions in the $s$-state, which to our knowledge cannot be 
obtained from Bethe ansatz computations, but still provides valuable physical 
information about the critical behaviour in the vicinity of $s=0$.
\medskip

Finally we remark that for the driven systems investigated here we have not 
been able to access the $s>0$ states which probe realisations of the process 
with a current {\em opposite} to the average current. 
This is likely because instead of remaining homogeneous in time and space, the 
typical trajectories develop heterogeneities in this situation. 
We may speculate that, as discussed by Bodineau and Derrida 
\cite{bodineauderrida2, bodineauderrida3}, non-stationary profiles, perhaps 
with shock-like spatial structures, will play an important role. 
This issue would certainly be worthwhile further detailed investigations.

\ack
This work has been supported in part by the U.S. National Science Foundation 
through grant NSF DMR-0308548 (UCT), and by the French Ministry of Education 
through the Agence Nationale de la Recherche's programme JCJC/CHEF 
(VL and FvW).
UCT gratefully acknowledges the warm hospitality at the Laboratoire de 
Physique Th\'eorique, Universit\'e de Paris-Sud, Orsay, France, where this work
was incepted and largely completed.
We thank John Cardy, Fabian Essler, Beate Schmittmann, Robin Stinchcombe, and 
Royce Zia for illuminating discussions.


\section*{References}

\begin{thebibliography}{10}

\bibitem{varadhan88}
  Varadhan S R S, {\it Large deviations and applications}, in
  {\it \'Ecole d'\'Et\'e de Probabilit\'es de Saint-Flour XV-XVII}, 
  ({\it Lecture Notes in Mathematics}, vol 1362)
  ed Hennequin P L (Berlin: Springer, 1988)

\bibitem{evanscohenmorriss}
  Evans D J, Cohen E G D, and Morriss G P, \PRL {\bf 71} 2401 (1993)

\bibitem{gallavotticohen}
  Gallavotti G and Cohen E G D, \PRL {\bf 74} 2694 (1995)

\bibitem{evanssearles94}
  Evans D J and Searles D J, \PR E {\bf 50} 1645 (1994)
  
\bibitem{jarzynski}
  Jarzynski C, \PRL {\bf 78}, 2690 (1997) 

\bibitem{kurchan}
  Kurchan J, \JPA {\bf 31} 3719 (1998)

\bibitem{lebowitzspohn}
  Lebowitz J L and Spohn H, {\it J. Stat. Phys.} {\bf 95} 333 (1999)

\bibitem{ldf-exp}
  Bustamante C, Liphardt J and Ritort F,
  {\it Physics Today} {\bf 58}, 43 (2005);
%
  Ritort F, 
  in {\it Poincar\'e Seminar} {\bf 2}, 195 (Birkh\"auser: Basel, 2003);
%
  Wang G M, Sevick E M, Mittag E, Searles D J, and Evans D J,
  \PRL {\bf 89} 050601 (2002);
%
  Wang G M, Reid J C, Carberry D M, Williams D R M, Sevick E M, and Evans D J,
  \PR E {\bf 71} 046142 (2005);
%
  Wang G M, Carberry D M, Reid J C, Sevick E M, and Evans D J, \JPCM
  \emph{J. Phys. Condens. Matter} {\bf 17} S3239 (2005);
%
  Feitosa K and Menon N,
  \PRL {\bf 92} 164301 (2004);
%
  Ciliberto S, Garnier N, Pinton J F, and Ruiz-Chavarria R,
  {\it Physica A} {\bf 340} 240 (2004);
%
  Ciliberto S and Laroche C, {\it J. Physique IV} {\bf 8} Pr6-215 (1998);
%
  Garnier N and Ciliberto S, \PR E {\bf 71} 060101 (2005);
%
  Ritort F, {\it J. Phys. Conden. Matter} {\bf 18} R531 (2006)

\bibitem{bodineauderrida1}
  Bodineau T and Derrida B, \PRL {\bf 92} 180601 (2004)

\bibitem{bodineauderrida2}
  Bodineau T and Derrida B, \PR E {\bf 72}, 066110 (2005)

\bibitem{bodineauderrida3}
  Bodineau T and Derrida B, {\it J. Stat. Phys.} {\bf 123} 277 (2006)

\bibitem{bertinidesole2}
  Bertini L, De~Sole A, Gabrielli D, Jona-Lasinio G, and Landim C, 
  \PRL {\bf 94} 030601 (2005)

\bibitem{bertinidesole3}
  Bertini L, De~Sole A, Gabrielli D, Jona-Lasinio G, and Landim C, 
  {\it J. Stat. Phys.} {\bf 123} 237 (2006)

\bibitem{bertinidesole4}
  Bertini L, De~Sole A, Gabrielli D, Jona-Lasinio G, and Landim C,
  e-print {\tt math.PR/0512394}

\bibitem{janssenschmittmann}
  Janssen H K and Schmittmann B, {\it Z. Phys. B -- Condensed Matter}
  {\bf 63} 517 (1986)

\bibitem{leungcardy}
  Leung K-t and Cardy J L, {\it J. Stat. Phys.} {\bf 44} 567 (1986)

\bibitem{forsternelsonstephen}
  Forster D, Nelson D R, and Stephen M J, \PR A {\bf 16} 732 (1977)

\bibitem{schmittmannzia1}
  Schmittmann B and Zia R K P,  
  {\it Statistical Mechanics of Driven Diffusive Systems}
  ({\it Phase Transitions and Critical Phenomena} vol~17) 
  ed Domb C and Lebowitz J L (London: Academic Press, 1995)

\bibitem{schmittmannzia2}
  Schmittmann and Zia R K P,  {\it Proc. IXth Int. Summer School on 
  Fundamental Problems in Statistical Mechanics} ed H van Beijeren 
  {\it Phys. Rep.} {\bf 301} 45 (1998)

\bibitem{honkonen}
  Honkonen J, \JPA {\bf 24} L1235 (1991)

\bibitem{bouchaudetal}
  Bouchaud J-P, Comtet A, Georges A, and Le~Doussal P, 
  \APNY {\bf 201} 285 (1990)

\bibitem{janssen1}
  Janssen H K, {\it Z. Phys. B} {\bf 23} 377 (1976)

\bibitem{dedominicis}
  De~Dominicis C, {\it J. Phys. (France)} Colloq. {\bf 37} C247 (1976)

\bibitem{bauschjanssenwagner}
  Bausch R, Janssen H K, and Wagner H, {\it Z. Phys. B} {\bf 24} 113 (1976)

\bibitem{janssen2}
  Janssen H K, {\it Dynamical critical phenomena and related topics} 
  ({\it Lecture notes in physics} vol 104) ed C P Enz (Berlin: Springer, 1979) 
  p 26

\bibitem{schaefer}
  Sch\"afer L, {\it Phys. Rep.} {\bf 301} 205 (1998)
  
\bibitem{taeuber}
  T\"auber U C, 
  \emph{Field theory approaches to nonequilibrium dynamics},
  in: ``Ageing and the Glass Transition'',
  {\it Proc. Int. Summer School Ageing and the Glass 
  Transition} ({\it Lecture notes in physics}) ed Henkel M, Pleimling M, and 
  Sanctuary R  (Berlin: Springer, 2007), Chap.~7, pp~295-348,
  e-print {\tt cond-mat/0511743},


\bibitem{derridalebowitz}
  Derrida B and Lebowitz J L, \PRL {\bf 80} 209 (1998)

\bibitem{vanbeijerenkutnerspohn}
  van~Beijeren H, Kutner R, and Spohn H, \PRL {\bf 54} 2026 (1985)

\bibitem{martinsiggiarose}
  Martin PC, Siggia ED, and Rose HA, {\em Phys. Rev. A} {\bf 8} 423 (1973)

\bibitem{andersen}
  Andersen HC, {\em J. Math. Phys.} {\bf 41} 1979 (2000)

\bibitem{giardinakurchanpeliti}
  Giardin\`a C, Kurchan J, and Peliti L, \PRL {\bf 96} 120603 (2006)

\bibitem{thermo-formalism}
  Lecomte V, Appert-Rolland C, and van Wijland F, 
  e-print {\tt cond-mat/0606211}, to appear in {\it J. Stat. Phys.}

\bibitem{derridaappert}
  Derrida B and Appert C, {\it J. Stat. Phys.} {\bf 94} 1 (1999) 

\bibitem{kim}
  Kim D, \PR E {\bf 52} 3512 (1995)

\bibitem{pathinteg-stoch}
  Suna A, \PR B {\bf 1} 1716 (1970);
  Doi M, \JPA {\bf 9} 1465 (1976);
  Peliti L, {\it J. Phys. (Paris)} {\bf 46} 1469 (1985)

\bibitem{vanwijlandracz}
  van~Wijland F and R\'acz Z, {\it J. Stat. Phys.} {\bf 118} 27 (2005);
  Lecomte V, R\'acz Z, and van Wijland F, {\it J. Stat. Mech.} P02008 (2005)

\bibitem{bouchaudetal87}
  Bouchaud J P, Comtet A, Georges A, and Le Doussal P,
  {\it J. Phys. (Paris)} \textbf{48} 1445 (1987);
  {\it J. Phys. (Paris)} \textbf{49} 369 (1988) 
\end{thebibliography}
\end{document}